\documentclass[%
 reprint,
 amsmath,amssymb,
 aps,
]{revtex4-2}
\PassOptionsToPackage{linktocpage}{hyperref}
\usepackage[hyperindex,breaklinks]{hyperref}
\usepackage{xcolor}
\usepackage{physics}
\usepackage{graphicx}% Include figure files
\usepackage{dcolumn}% Align table columns on decimal point
\usepackage{bm}% bold math
\usepackage{natbib}
\begin{document}

\preprint{APS/123-QED}

\title{Harnessing Quantum Dynamics for Robust and Scalable Quantum Extreme Learning Machines}% Force line breaks with \\

\author{Payal D. Solanki}
\email{paysolanki@deloitte.com}
\affiliation{Deloitte Consulting India Private Limited}
 
\author{Anh Pham}
 \email{anhdpham@deloitte.com}
\affiliation{Deloitte Consulting LLP}

\date{\today}% It is always \today, today,

\begin{abstract}
Quantum Extreme Learning Machine (QELM) is an emerging hybrid quantum machine learning framework that leverages quantum system dynamics to enhance classical models. However, QELM can suffer from the exponential concentration problem, where excessive entanglement reduces model expressivity. In this work, we gain insight into this challenge and demonstrate how tensor network methods specifically, the Time Dependent Variational Principle (TDVP) with Matrix Product States (MPS) can efficiently simulate quantum systems while controlling entanglement and mitigating exponential concentration. Using numerical experiments on the Modified National Institute of Standards and Technology (MNIST) dataset, we show that time-evolving an MPS system modeled as a chain of Rydberg atoms produces high-quality data embeddings with low classical computational overhead. Our findings indicate that exact simulation of quantum dynamics is not necessary for strong machine learning performance; even approximate quantum embeddings can yield competitive results. Furthermore, we observe that both increased disorder in the quantum state achieved by tuning Hamiltonian parameters and careful control of entanglement directly correlate with improved model accuracy, highlighting the importance of these factors in optimizing QELM performance.
\end{abstract}

\maketitle

%\tableofcontents

\section{Introduction}
The intersection of quantum computing and machine learning (ML) has catalyzed the emergence of quantum machine learning (QML), a rapidly evolving field that seeks to harness quantum phenomena to enhance classical data processing and pattern recognition tasks \cite{qml1,qml2,qml3,qml4,qml5}. Among the varied approaches within QML, hybrid frameworks that integrate quantum system dynamics with classical models such as Quantum Reservoir Computing (QRC) and Quantum Extreme Learning Machines (QELM) \cite{qrc0,qrc1,qrc2,qrc3,qrc4,qelm1,qelm2} have shown particular promise. These methods leverage the complex, high-dimensional dynamics of quantum systems to transform input data into expressive feature spaces, potentially enabling efficient solutions to problems that are intractable for classical algorithms.

Inspired by classical reservoir computing (RC) and Extreme Learning Machines (ELM)\cite{rc1, rc2, rc3,rc4, elm1, elm2, elm3}, QRC and QELM leverage the intrinsic dynamics of quantum systems to process data. In classical RC and ELM, a fixed dynamical system, known as the reservoir, transforms input data into a high-dimensional space, thus enhancing the features used for training through simple linear regression. Although RC and ELM seemingly similar, a main difference resides in the fact that RC exploits the natural dynamics of the substrate (reservoir) as an internal memory of past input information while ELM does not \cite{qrc4}. This is why RC is reserved for time series prediction, whereas ELM is used for tasks such as classification and regression. QRC and QELM extend this concept into the quantum realm, utilizing the exponentially large Hilbert space of quantum systems to generate correlations that are classically intractable. Recent experimental study of the QRC algorithm has shown that this QML method can be implemented effectively on analog quantum hardware up to 108 qubits by encoding classical data into different geometrical structures of Rydberg atoms \cite{qrc0}. Beyond Rydberg Hamiltonians, many studies have investigated the interplay between the quantum encoding layer and the Hamiltonian used to generate the quantum reservoir, and how this interaction can influence subsequent ML tasks. \cite{qelm_app,qelm_app_1, qelm_app_2, qelm_app_3, qelm_app_4,qelm_app_5}.

Although QELM shows great potential, recent literature points out major challenges, notably the risk of exponential concentration. This issue arises when expectation value measurements cluster around specific values regardless of the input, resulting in nearly indistinguishable feature representations and ultimately impairing the effectiveness of downstream classical ML algorithms. Such a phenomenon is intrinsically linked to excessive entanglement, originated from the quantum feature encoding scheme or the Hamiltonian evolution generating the quantum reservoir \cite{qelm1}. As a result, understanding the physics to carefully manage the entanglement growth within the quantum dynamics is essential for optimal performance in QELM.

Among many quantum many-body systems, the 1D Rydberg model exhibits rich quantum phases by tuning internal parameters within the Hamiltonian such as the Rabi frequency, detuning, as well the distance between the atoms within the chain \cite{1D_rydberg_physics, 1D_rydberg_physics_2, 1D_rydberg_physics_3, 1D_rydberg_physics_4}. In addition, QRC and QELM have been shown to be viable on system like Rydberg based quantum computer \cite{qrc0, qrc5, qrc6}. Within this study, we leverage the quantum physics of the 1D Rydberg chain to optimize the performance of our QELM model by tuning the different parameters within the Hamiltonian to produce optimal quantum features. 

%The promise of many QML algorithms lies in the ability to use near-term quantum devices to produce correlations that would otherwise require exponential classical resources to simulate \cite{qml1}. 

Beyond exact simulation of quantum dynamics, quantum-inspired algorithms based on tensor network (TN) methods have emerged as an important tool to study quantum many-body systems including Rydberg-based models  \cite{tn1,tn2,tn3,tn4,tn5, tn8, tn9, tn10}. TN algorithms offer a key advantage: they efficiently represent and manipulate quantum states with limited entanglement, enabling the simulation of large quantum systems on classical hardware by compressing the state space \cite{tn6, tn7, tn12}. Specifically, TNs can efficiently simulate quantum state evolutions \cite{tn6, dyn1, dyn2, dyn3, dyn4, tn11} and compute expectation values \cite{tn2, tn12}  for certain quantum systems within a certain approximation of their dynamics \cite{tn14, dyn5}. Secondly, TNs scale favorably with an increasing number of qubits for systems with low entanglement, mitigating the exponential complexity by exploiting quantum correlations, thus allowing the simulation of large-scale quantum systems with reduced computational overhead \cite{tn1}. Finally, since our quantum reservoir is generated by evolution over time, the nonequilibrium state after the evolution can result in entanglement spreading throughout the system \cite{dyn10,dyn11}, resulting in lower performance of the ML model due to the problem of exponential concentration. Consequently, compression using TNs not only reduces computational resources, but also provides a means of controlling the entanglement structure which is particularly relevant for addressing the exponential concentration problem observed in QELM \cite{qelm1}. Thus, TN methods present a practical and scalable pathway to advance QELM and other quantum-inspired machine learning frameworks on classical devices.

In this paper, we leveraged TN algorithms, specifically the time-dependent variational principle (TDVP) \cite{dyn2, dyn3, dyn4} and the matrix product state (MPS) to perform the time-evolution of 1D Rydberg atoms, with the goal of providing an example of how the QELM technique can be optimized and implemented at scale using a quantum-inspired technique. MPS-TDVP techniques based on the Lie-Trotter decomposition have been known to achieve a balance between performance and accuracy when they are used to simulate quantum dynamics such as time evolution \cite{dyn4}. Specifically, TDVP restricts dynamics of the many-body wavefunction within a variational MPS manifold during the time evolution \cite{dyn5}. As a result, this can provide a pathway to manage the entanglement growth during the time evolution. To demonstrate the efficacy of TNs, we performed numerical experiments using the MNIST \cite{mnist} dataset using our MPS-based QELM for classification. Our results suggest that the TDVP and MPS techniques are highly scalable, which opens up more possibilities to apply QELM to datasets with more features using classical computational resources. 

In short, our experiments show that the TDVP-MPS approach enables efficient simulation of quantum dynamics for systems with a higher number of qubits on classical hardware, with computational overhead that scales favorably. We find that achieving strong ML performance does not require highly precise simulation of quantum dynamics; even approximate embeddings generated by scalable TN algorithms yield classification accuracies comparable to those of nonlinear classical models, including neural networks (NNs). Furthermore, the degree of disorder in the quantum state controlled by tuning Hamiltonian parameters directly correlates with improved model performance. Increased disorder enhances the diversity and expressiveness of the quantum embeddings, leading to higher classification accuracy. Analysis of quantum entanglement reveals that, while some entanglement is necessary to enrich the feature space, excessive entanglement can result in feature concentration and diminished model performance.
\begin{figure*}[t]
%\centering \includegraphics[scale=0.52]{qelm.png} 
\centering \includegraphics[scale=0.515]{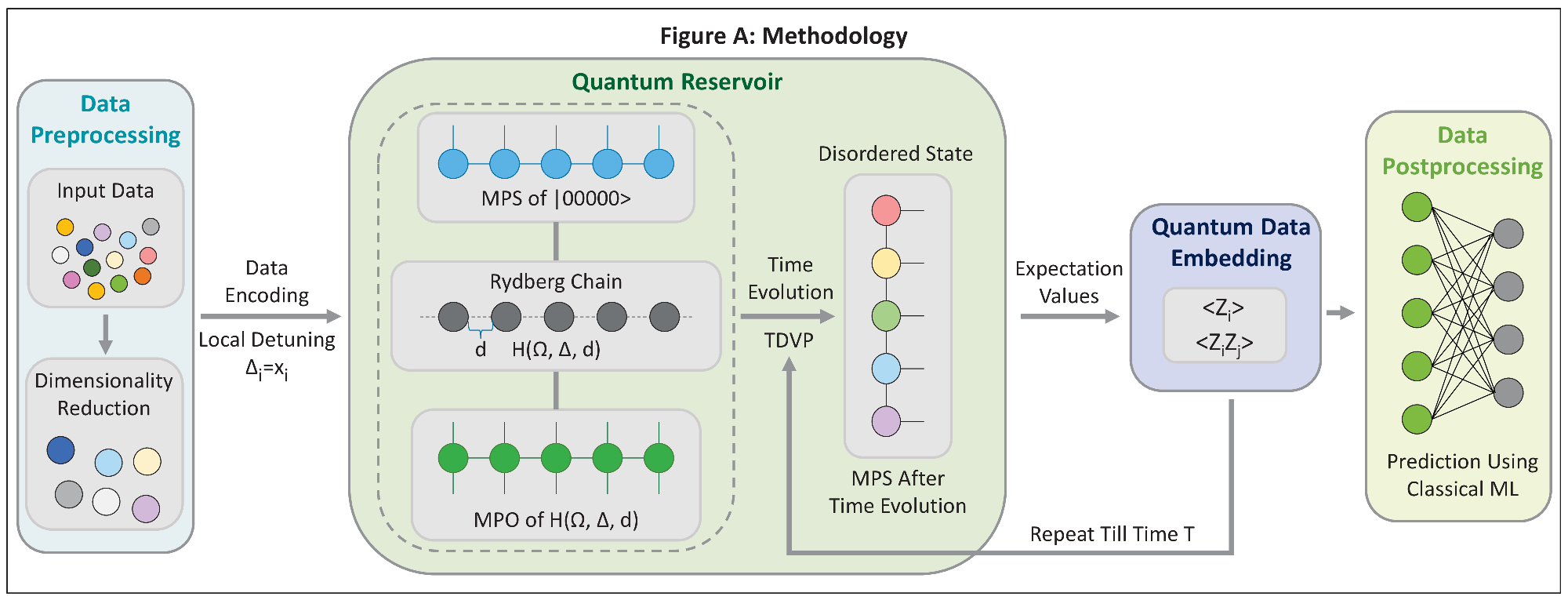}
\centering \includegraphics[scale=0.63]{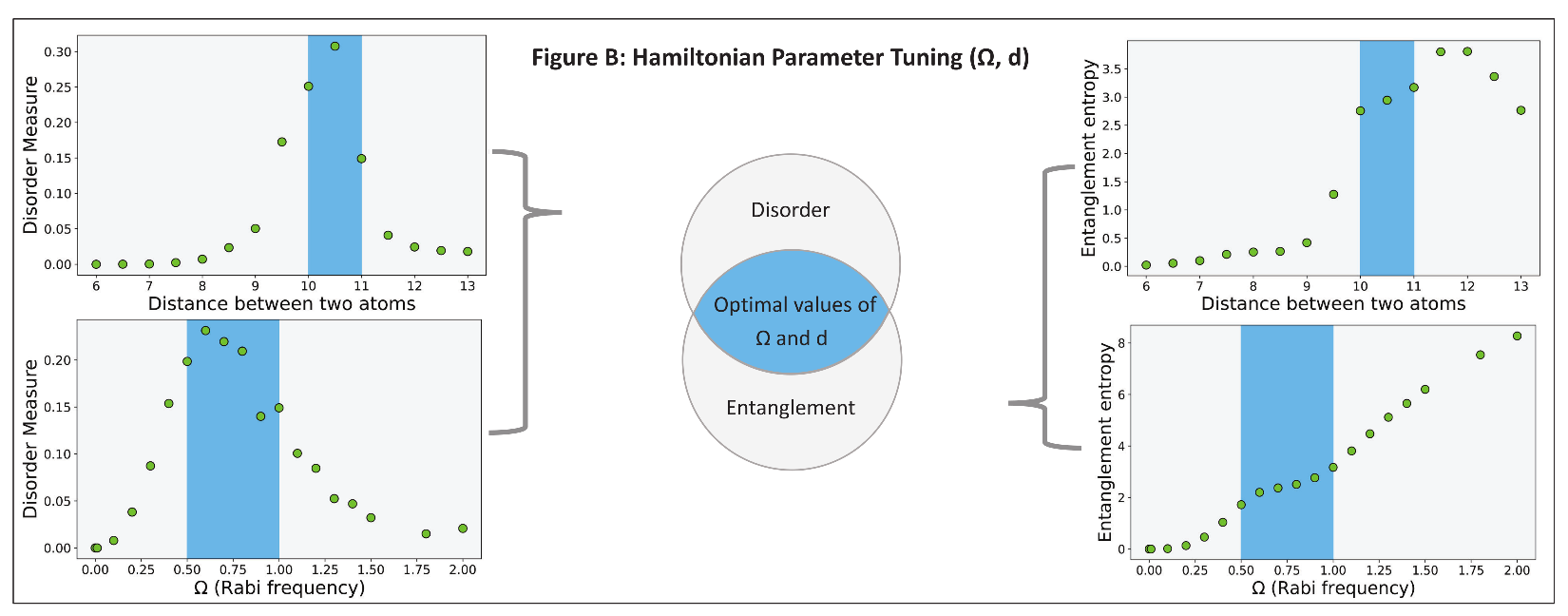}
\caption{
A schematic representation of our tensor network approach for QELM with Rydberg Hamiltoninan embedding and how the different parameters in the  Hamiltonian can be used to control entanglement and disorder in the Rydberg dynamics. (a) The workflow includes data preprocessing, constructing the embedding by computing the expectation values of correlators with tensor network formalism at each time step $dt$ up to total time $T$, and finally performing prediction using classical machine learning regression methods. (b) The figure also demonstrates the optimization of Hamiltonian parameters, such as the Rabi frequency ($\Omega$) and the distance between two atoms. The highlighted regions represent parameters which yield optimal ML performance. By fixing one parameter and varying the other, the system's disorder can be maximized, which enhances machine learning accuracy as greater disorder generally leads to better performance. Additionally, the effect of entanglement entropy is shown, indicating that a moderate amount of entanglement is optimal for the algorithm; both too little and too much entanglement can be detrimental. This data was generated using the TDVP two-site algorithm.
}
\label{qelm_pic}
\end{figure*}
\section{Methodology}
A schematic workflow of our TN simulation of Rydberg Hamiltonian dynamics to generate the new embedding for ML task is shown in Fig.(\ref{qelm_pic}a). In general, a QELM algorithm consists of three primary components: data encoding using quantum feature map, performing Hamiltonian simulation with time evolution to increase the feature space; obtaining a quantum embedding through expectation value calculation; and training a simple classical ML model using the new quantum embedding. In our approach, we utilized a Rydberg Hamiltonian and its dynamics to encode the data as well as to generate the new quantum embeddings. In addition, rather than performing exact simulation of the quantum dynamics, TN methods like MPS and TDVP methods were used to perform the encoding and time-evolution in a 1D Rydberg chain.
\subsection{Rydberg Hamiltonian Encoding and Quantum Embedding generation}\label{QRC_method}
 %The first step involves preprocessing the data to make it suitable for supervised learning tasks. The training and testing datasets consist of data pairs, ${(x_i[n], y_j[n])}$, where $x_i[n]$ are the input feature vectors and $y_j[n]$ are the corresponding labels. In this notation, $n$ indexes the individual data points, while $i$ and $j$ index the components of the feature and label vectors, respectively.
This method begins by encoding the data features into the parameters of the Rydberg Hamiltonian \cite{rb1}, as described in:
\begin{align}
H = \sum_{j} \left[ \frac{\Omega_j}{2} \left( e^{i \phi_j} \left| g_j \right\rangle \left\langle r_j \right| + e^{-i \phi_j} \left| r_j \right\rangle \left\langle g_j \right| \right) \right]\nonumber\\ - \sum_{j} \Delta_j \hat{n}^j + \sum_{j<k} V_{jk} \hat{n}^j \hat{n}^k
\end{align}

In this Hamiltonian, $\Omega_j$ represents the Rabi drive amplitude between a ground state ($\ket{g_j}$, with $j$ indexing the atoms) and a highly excited Rydberg state ($\ket{r_j}$). $\Delta_j$, the detuning of the driving laser field for atom $j$, and $\phi_j$, the laser phase for atom $j$. Furthermore, $V_{jk}$ describes the van der Waals interaction between atoms $j$ and $k$, which can be derived from the geometry of the lattice as $V_{jk}=C/\lVert\mathbf{r}_j-\mathbf{r}_k\rVert^6$. The value of $C$ is $862690 \times 2\pi$ MHz$\cdot\mu$m$^6$. In our study, the atoms are arranged in a one-dimensional chain with equal spacing $d$ between each atom.

To express the Hamiltonian in terms of Pauli matrices, we can use the following substitutions: $| g_j \rangle \langle r_j |$ is replaced by $\sigma_j^- = (\sigma_j^X - \dot{\iota} \sigma_j^Y)/2$, which is the lowering operator; $| r_j \rangle \langle g_j |$ is replaced by $\sigma_j^+ =( \sigma_j^X + \dot{\iota} \sigma_j^Y)/2$, which is the raising operator; and $\hat{n}^j = | r_j \rangle \langle r_j |$ is replaced by $(1 + \sigma_j^Z)/2$, which is the number operator.
Using these substitutions, the Hamiltonian can be written as:
%\begin{widetext}
\begin{eqnarray}
H &=&\sum_{j} \left[ \frac{\Omega_j}{2} \left( \cos \phi_j ~\sigma_j^X - \sin \phi_j ~\sigma_j^Y \right) \right]\nonumber\\&&- \sum_{j} \frac{\Delta_j}{2} \left(1+\sigma_j^Z\right) + \sum_{j<k}\frac{V_{jk}}{4} \left(1+\sigma_j^Z+\sigma_k^Z+\sigma_j^Z\sigma_k^Z\right)\nonumber\\
\label{ham1}
\end{eqnarray}
%\end{widetext}
Data encoding is achieved through site-dependent local detunings, represented as $\Delta_j =  x_j$ \cite{qrc0}. Consequently, a $N$-qubit system is capable of encoding $N$ features.  Once the detuning values are fixed through data encoding, the remaining Hamiltonian parameters to set are the Rabi frequency $\Omega$ and the interatomic distance $d$, which determines the interaction strength $V_{jk}$.The selection and influence of these parameters are discussed in detail in the Results and Discussion section. As illustrated in Fig.(\ref{qelm_pic}b), optimal model performance is achieved by tuning the Hamiltonian parameters to maximize disorder in the quantum states while maintaining moderate entanglement.
Following data encoding, the final step involves classical post-processing. To achieve this, we obtain the data embedding vectors, which can be derived from the Hamiltonian dynamics of the quantum system. Specifically, after encoding the data through site-dependent local detunings, represented as $\Delta_j = x_j$, the $N$-qubit system transitions from an all spin-up ground state under the influence of the specifically designed Rydberg Hamiltonian. The quantum dynamics are then examined over several successive time steps. At each time step, the expectation values of local observables are measured, typically one- and two-point correlators on a computational basis, such as $\langle Z_j \rangle$ and $\langle Z_j Z_k \rangle$. These local observables then form the data embedding vectors, $u_i[n]$, with $i$ indexing the different correlators and probe times. These vectors are essential for the classical post-processing step. To obtain these expectation values, the TN algorithm has been used.
\subsection{Tensor Network Methods for Dynamical Simulation of Rydberg Hamiltonian for QELM}\label{TN_method}
In the realm of TN algorithms, the TDVP is a robust method to simulate the dynamics of quantum systems, particularly those with a substantial number of qubits. TDVP can be implemented in two primary forms: single-site TDVP and two-site TDVP. Both approaches are crucial for efficiently evolving quantum states represented by MPS under the influence of a Hamiltonian, represented by the Matrix Product Operator (MPO) format.
Single-site TDVP focuses on updating one site (or tensor) of the MPS at a time, achieving  relatively low computational cost and suitability for large systems while maintaining unitary time evolution, energy conservation \cite{dyn3}, and numerical stability \cite{dyn6, dyn7}. However, single-site TDVP uses a fixed-rank integration scheme, which does not allow the bond dimension to grow during the time evolution, limiting its applications to simulate only certain quantum dynamics \cite{dyn5, dyn8}. In contrast, TDVP for two sites updates pairs of neighboring sites simultaneously, allowing for a more accurate representation of correlations, especially in strongly interacting systems, but has a much higher computational cost \cite{dyn2}. 

For the simulation of the quantum system in this study, the TeNPy library \cite{tenpy} was utilized. TeNPy is a versatile and efficient library for implementing TN algorithms, including single-site and two-site TDVP. It provides robust tools for handling MPS and MPO representations, making it ideal for simulating quantum system evolution. In this simulation, we encoded the classical data in a 1D chain. The Rydberg Hamiltonian was represented using the MPO format, and the quantum system, initially set in an all-up state,was evolved using the TDVP algorithm. Correlators were computed at each time step to generate the nonlinear embedding.

It is important to note that the interaction term $V_{jk}$ decreases very rapidly as the distance between the atoms increases. Consequently, $ V_{jk}$ becomes negligible for the farthest neighbors, which can be safely ignored in the calculations. This rapid decrease in interaction strength allows us to reduce the dimensions of MPOs, leading to a significant reduction in computational time. Furthermore, we can optimize data embedding by excluding expectation values $\langle Z_j Z_k \rangle$ for pairs of sites $j$ and $k$ that do not interact in the Hamiltonian. By focusing only on the relevant interactions, we can streamline our data processing and further enhance computational efficiency. For instance, if we consider the case of 25 qubits, the original number of correlators for a particular time step would be 325. However, by choosing a truncation limit for $V_{jk}$ of $10^{-4}$, the number of correlators can be reduced to 247. The detailed analysis of $V_{jk}$ threshold is given in \cite{supplemental}.

\subsection{Data Preprocessing, QRC Simulation, and Classical Machine Learning Methods}\label{sectionc}

To investigate the efficacy of our quantum-inspired QRC technique, we applied the method to perform a classification task using the MNIST dataset \cite{mnist}. The initial step in our methodology involves preprocessing the MNIST dataset, which consists of 60,000 training samples and 10,000 test samples, each represented as grayscale images of $28\times 28$ pixels. Principal Component Analysis (PCA) \cite{pca} was used to reduce the dimensionality of the data to facilitate the embedding process. Additionally, we rescaled the $x_j$ inputs to the range $[-6, 6]$. Each component of the PCA-reduced data was mapped to a qubit as described in the methodology. 
\begin{figure}[b]
    \centering
    \includegraphics[scale=0.33]{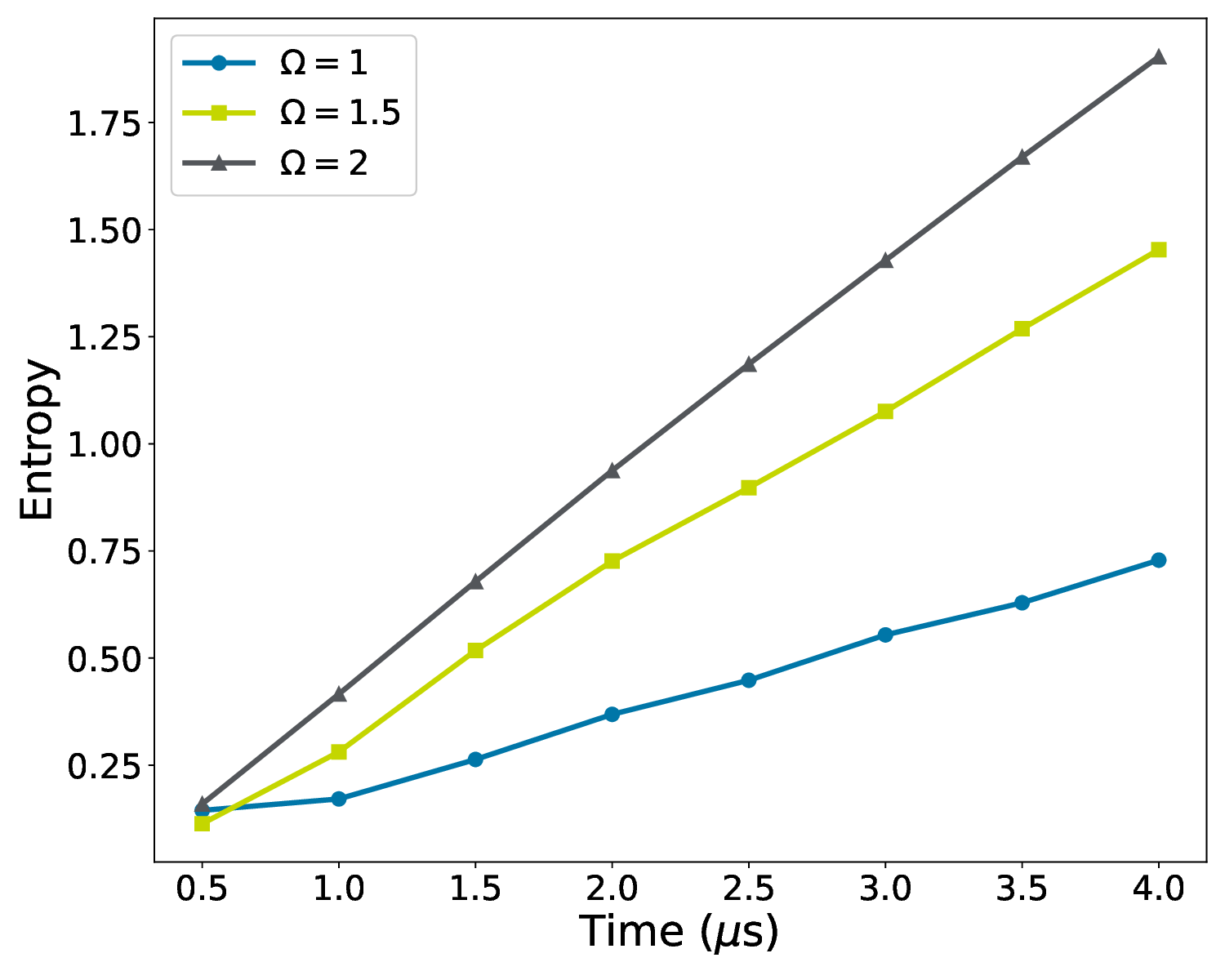}
    \caption{This plot illustrates the variation of entanglement entropy with increasing time for different values of $\Omega$. This plot was generated using TDVP two site algorithm.}
    \label{entropy}
\end{figure}
The embeddings were generated by computing the expectation values of the local observables at each time step. Throught this paper, the simulation was run with a time step of $0.5 \, \mathrm{\mu s}$ up to a total time of $4 \, \mathrm{\mu s}$. A detailed analysis of the effect of total evolution time on model performance is provided in \cite{supplemental}.
\begin{figure*}[t]
    \centering
    \includegraphics[scale=0.37]{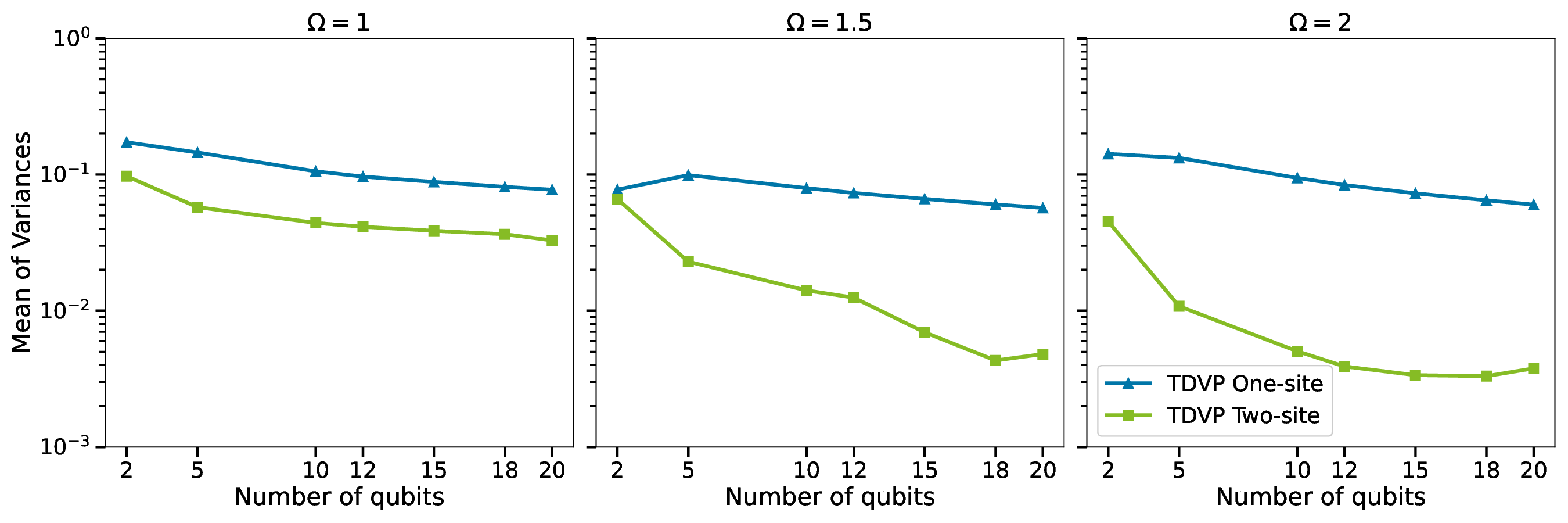}
    \caption{This plot illustrates the exponential concentration of correlators as a function of system size for varied values of $\Omega$ around the optimal region. Shown is the mean of variance of correlators across the dataset as a measure of concentration. Results are compared for both TDVP one-site and TDVP two-site methods, highlighting the stronger concentration effect observed in the two-site approach.}
    \label{concen}
\end{figure*}

Once the quantum embeddings were obtained, we performed classification using a simple linear fitting model with the quantum embeddings as input. In addition, we also compared the performance of our quantum-inspired QELM model with a linear and nonlinear ML models using the original PCA-reduced data as input. All of our ML classification tasks were performed using the keras package \cite{keras}. The models and their parameters are described in \cite{supplemental}. The accuracy of the models was determined by comparing the predicted labels with the true labels of the test set, enabling a direct comparison of the performance between the quantum embeddings and the PCA embeddings.

In this investigation, we evaluated the performance of the quantum-inspired QELM model using various hyperparameters. For the machine learning process, we utilized 10 features and a training set comprising 10,000 data points and a test set consisting of 1,000 data points. Additionally, we employed $k$-fold cross-validation with $k=5$ to ensure robust evaluation. The standard deviation across the folds are represented as error bars in all figures.

\begin{figure*}
    \centering
    \includegraphics[scale=0.35]{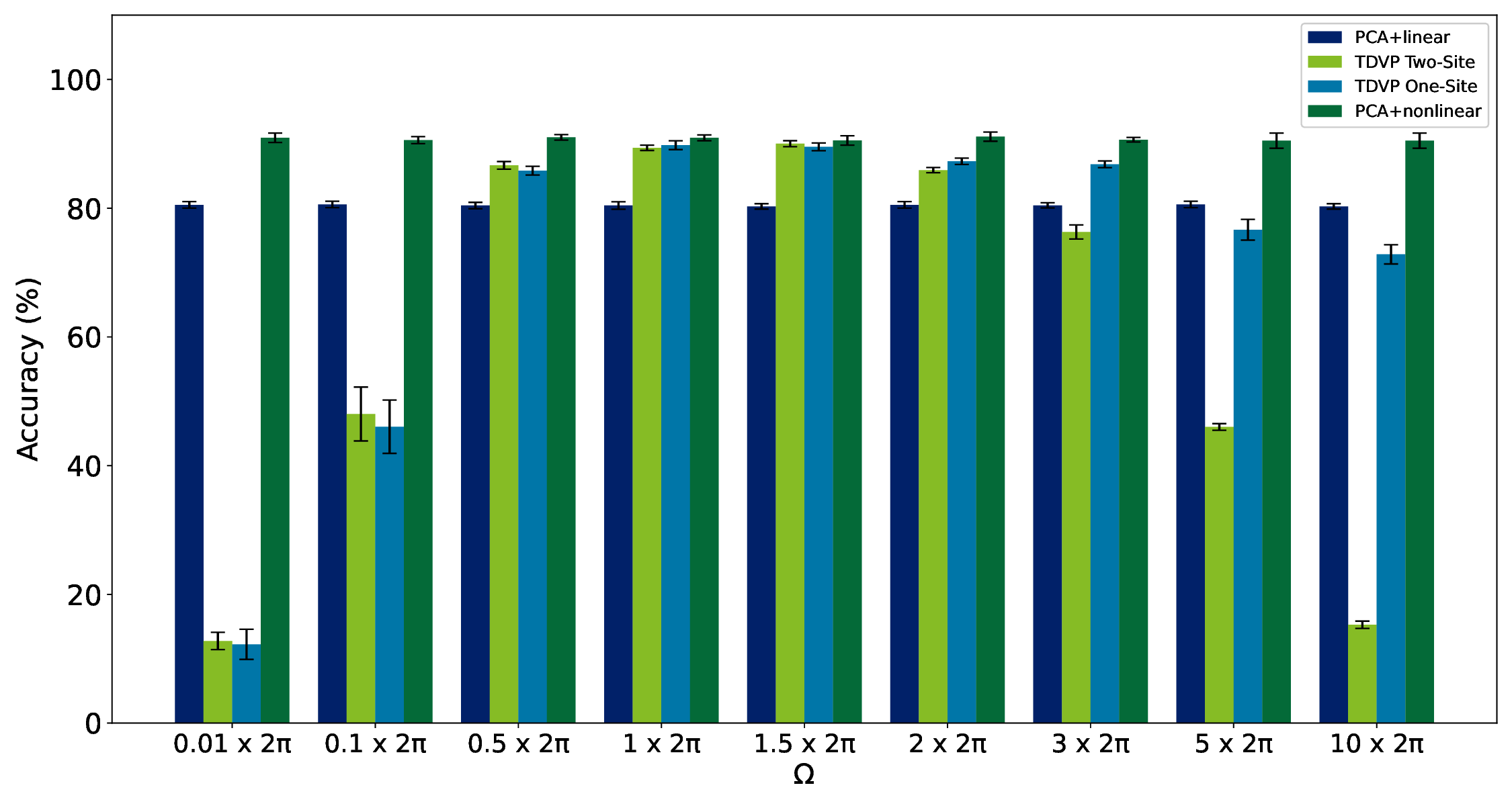}
    \caption{\textbf{Quantum inspired QELM performance for 10 qubits (10 features) with varying $\Omega$}. This result was compared against the quantum embedding generated by the TDVP one site and TDVP two site method. For this study, we also compared the quantum-inspired results against results from classical models (linear and NN) with classical features.}
    \label{entropy_vs_omega}
\end{figure*}
\begin{figure*}
    \centering
    \includegraphics[scale=0.4]{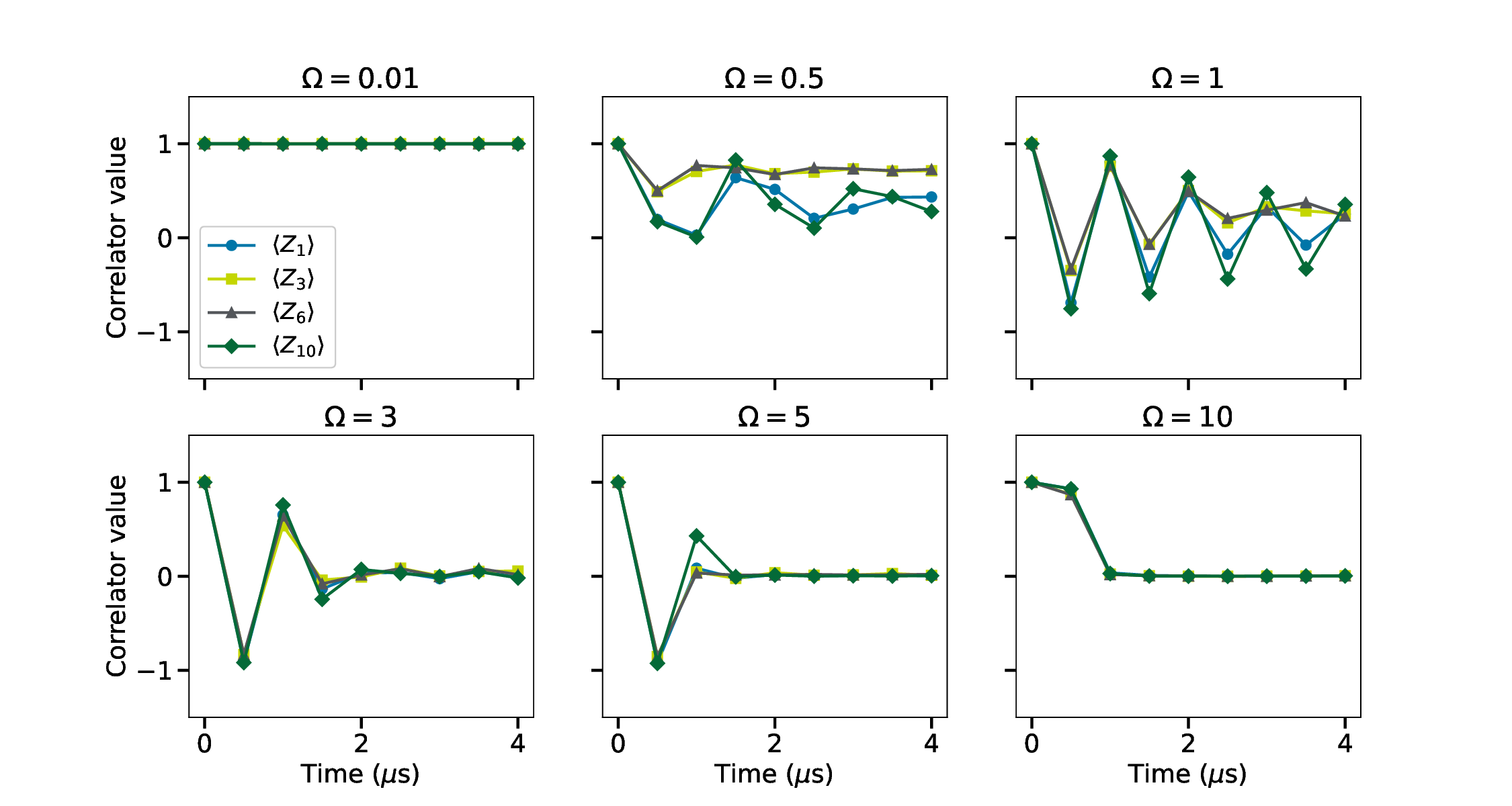}
    \caption{This figure illustrates the time evolution of various one-body correlators for different values of $\Omega$., generated using the TDVP two site algorithm}
    \label{steady_state}
\end{figure*}

\section{Results and Discussions}
A central finding of our investigation is that robust and scalable quantum-inspired QELM can be realized by leveraging the interplay between disorder and entanglement in quantum dynamics, and that highly accurate quantum simulation is not necessary to achieve strong ML performance. As illustrated in Fig.(\ref{qelm_pic}b), tuning the Hamiltonian parameters, specifically the Rabi frequency ($\Omega$) and interatomic distance ($d$), enables direct control over the degree of disorder and entanglement in the system, which in turn governs the expressiveness of the quantum embeddings and the resulting model accuracy. In the following, we systematically analyze how disorder and entanglement affect QELM performance, and why the TDVP one-site TN method emerges as the optimal approach for scalable quantum-inspired ML. We conclude with a discussion of scalability and the impact of feature dimensionality.

\subsection{Entanglement and  Concentration}
A central aspect of our study is understanding how entanglement, driven by the Rabi frequency $\Omega$, shapes the expressiveness of quantum embeddings and, ultimately, the performance of QELM models. We analyze this relationship through the lens of entanglement entropy, feature concentration, and model accuracy.
We begin by examining the growth of entanglement entropy as a function of time and $\Omega$, as shown in Fig.(\ref{entropy}). For small values of $\Omega$, the system remains nearly separable, and the entanglement entropy stays low throughout the evolution. As $\Omega$ increases, the entanglement entropy rises more rapidly, reflecting the enhanced quantum correlations generated by the Rydberg Hamiltonian. This trend is especially pronounced for larger $\Omega$, where the system quickly develops significant entanglement during time evolution. Thus, $\Omega$ acts as a direct control knob for the degree of entanglement in the system.
The consequences of growing entanglement are further revealed by examining the phenomenon of exponential concentration, depicted in Fig.(\ref{concen}). Here, we plot the variance of the single-site correlators $\langle Z_i \rangle$ across the dataset as a function of the number of qubits for different values of $\Omega$. For the TDVP two-site method, we observe that as either the system size or $\Omega$ increases, the variance of the correlators decreases exponentially. This exponential concentration of correlators indicates that, with high entanglement, the quantum embeddings for different input records become increasingly similar effectively “collapsing” the feature space and reducing the model’s ability to distinguish between inputs. In contrast, the TDVP one-site method exhibits a much weaker concentration effect, preserving greater diversity in the embeddings even as the number of qubits grows.
The impact of these effects on model performance is summarized in Fig.(\ref{entropy_vs_omega}), which shows the classification accuracy of QELM models (using both TDVP one-site and two-site methods) and classical baselines as a function of $\Omega$. For small $\Omega$, the system’s dynamics are limited, and accuracy is low. As $\Omega$ increases, accuracy improves and reaches a maximum at an intermediate, optimal value of $\Omega$ the “sweet spot” where entanglement is sufficient to enrich the feature space but not so large as to induce exponential concentration. Beyond this point, further increases in $\Omega$ cause accuracy to decline, especially for the TDVP two-site method, as excessive entanglement leads to rapid feature space concentration and loss of discriminative power. The TDVP one-site method, by discounting the entanglement, avoids this detrimental effect and maintains higher accuracy at large $\Omega$.

To further elucidate the underlying dynamics, we analyze the time evolution of the single-site correlators $\langle Z_i \rangle$ for different $\Omega$ which is shown in Fig.(\ref{steady_state}). At small $\Omega$, the correlators exhibit minimal oscillatory behavior and remain close to their initial values, indicating limited dynamical evolution. As $\Omega$ increases, the correlators display pronounced oscillations, typically spanning the full range between $-1$ and $1$. For large $\Omega$, these oscillations are rapidly damped, and the system quickly relaxes to a steady state. This accelerated relaxation at high $\Omega$ further reduces the diversity of transient features available for learning, contributing to the observed decline in model accuracy.

Overall, our results demonstrate that moderate entanglement controlled by $\Omega$ is essential for generating expressive quantum embeddings, but excessive entanglement leads to exponential concentration and degraded model performance. The TDVP one-site method, by discounting entanglement growth, preserves feature diversity and enables robust and scalable QELM, especially in regimes where the TDVP two-site method suffers from feature space collapse.

\begin{figure}[t]
   \centering
   \includegraphics[scale=0.295]{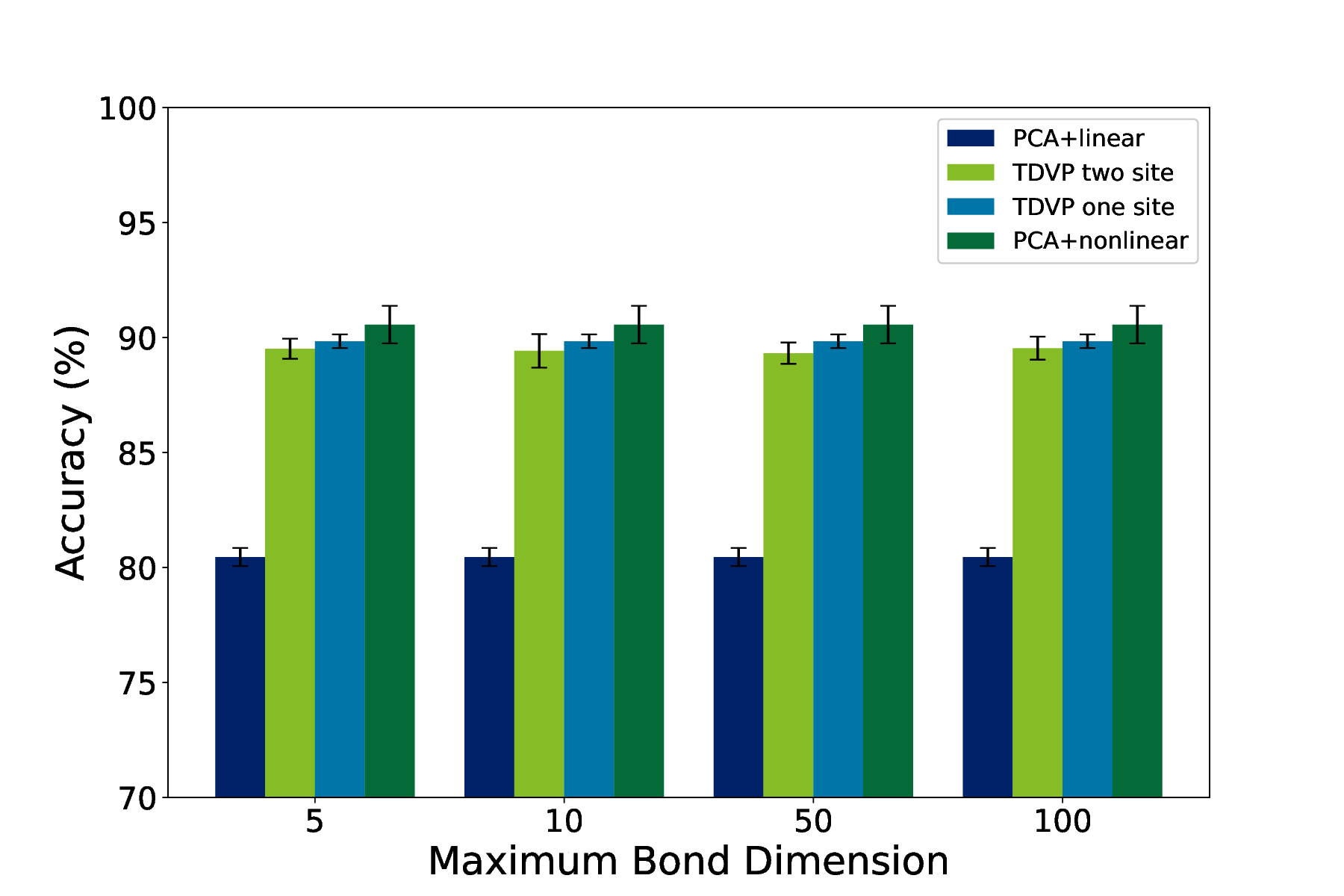}
   \caption{\textbf{Quantum inspired QELM performance for 10 qubits (10 features) with varying bond dimensions with TDVP two site algorithm}. Here the quantum embeddings were generated with the two-site TDVP method at different bond lengths. This result was compared against the quantum embedding generated by the TDVP one site method. For this study, we also compared the quantum-inspired results against results from classical models (linear and NN) with classical features.}
   \label{bond1}
\end{figure}

Subsequently, to further investigate the role of entanglement on our QELM performance, we examined the effect of correlator accuracy and bond dimension in the TN simulations. While the TDVP two-site algorithm yields correlators that are significantly more accurate than those from the TDVP one-site method for more details see \cite{supplemental}, we find that this increased accuracy does not translate into improved model performance. As illustrated in Fig.(\ref{bond1}), varying the bond dimension in the TDVP two site method thereby increasing the precision of the quantum simulation has negligible effect on the classification accuracy of the QELM model.
This observation highlights a central finding: for QELM applications, the precise accuracy of the quantum correlators is not critical. Even approximate embeddings generated by the efficient TDVP one-site method are sufficient to achieve strong model performance. This robustness further underscores the practicality and scalability of quantum-inspired approaches for ML, where computational efficiency can be prioritized without sacrificing model accuracy.
Interestingly, the TDVP one-site method generates quantum states with zero entanglement, yet still achieves model accuracy comparable to the entangled states produced by the TDVP two-site method. This suggests that, beyond entanglement, another phenomenon must be responsible for the expressiveness of the quantum embeddings and the observed model performance. As we show in the following section, disorder in the quantum dynamics emerges as a key factor enabling robust and effective QELM.

\begin{figure*}[t]
    \centering
    \includegraphics[scale=0.3]{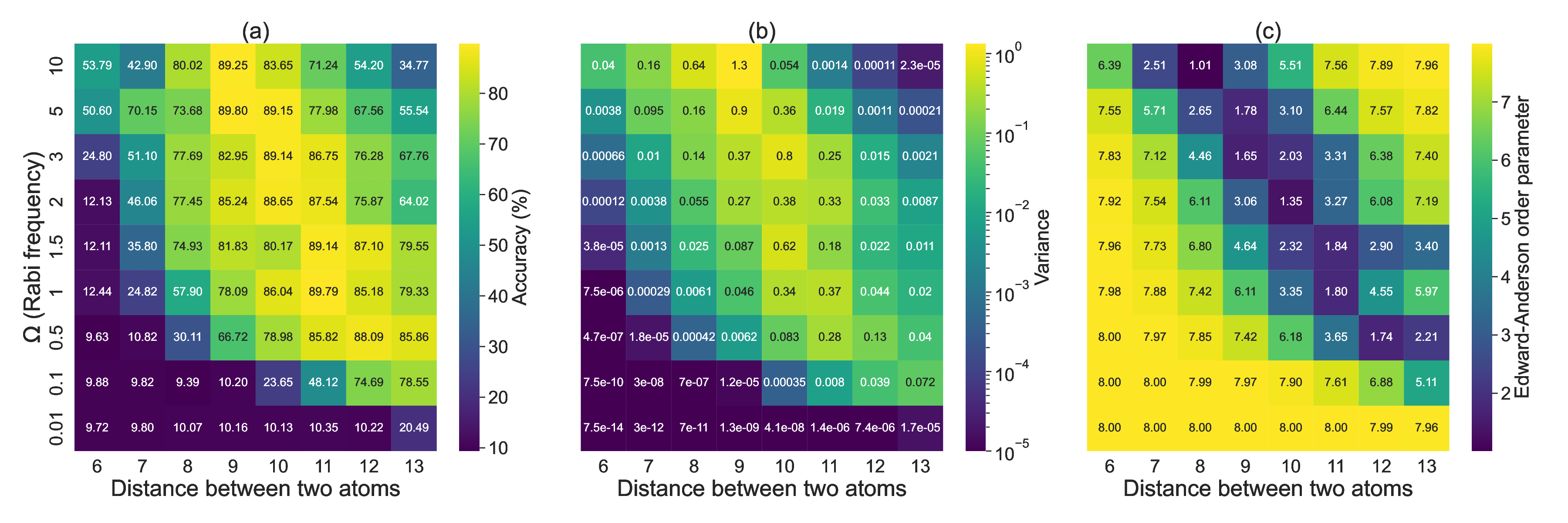}
   \caption{{\bf Analysis of Disorder Effects on Model Performance}
These heatmaps collectively illustrate how disorder influences model performance as a function of Hamiltonian parameters: Rabi frequency ($\Omega$) and the distance between nearest-neighbor atoms. All $\Omega$ values are expressed as multiples of $2\pi$, and embeddings are generated using the TDVP one-site algorithm. Disorder is quantified using both the variance of the correlator $\langle Z_i \rangle$ across sites and the Edwards-Anderson order parameter.
(a) Machine learning model accuracy.
(b) Variance of $\langle Z_i\rangle$ across sites, summed over all time steps.
(c) Edwards-Anderson order parameter, summed over all time steps.}
    \label{perf_heatmap}
\end{figure*}
\subsection{Disorder and Model Expressiveness}
Disorder in the quantum system, controlled by the Hamiltonian parameters, emerges as a crucial factor in determining the expressiveness of quantum embeddings and the resulting ML performance. In our QELM framework, disorder arises from the interplay between the Rabi frequency $\Omega$; the interatomic interaction strength $V_{jk}$; which is set by the distance between atoms, and the site-dependent detuning $\Delta_i$; which is fixed by the scaled input features.
To systematically investigate the impact of disorder, we varied both $\Omega$ and the interatomic distance while keeping $\Delta_i$ determined by the input data. The resulting QELM model accuracy is shown as a heatmap in Fig.(\ref{perf_heatmap}a). Notably, the model accuracy exhibits a non-monotonic dependence on both parameters: accuracy initially increases with $\Omega$ and distance, reaches an optimal value and then declines as either parameter becomes too large. This demonstrates that there exists an optimal regime in parameter space where the QELM model achieves its highest accuracy. Specifically, peak performance is observed at intermediate values of both $\Omega$ and the interatomic distance, while values that are too low or too high lead to diminished accuracy.
One of the central findings of our study is that the degree of disorder in the quantum state, as tuned by these Hamiltonian parameters, has a direct and significant impact on ML performance. Increased disorder, resulting from the competition between the Rabi drive, pairwise interactions, and inhomogeneous detuning, leads to richer quantum dynamics and more diverse quantum embeddings. This, in turn, enhances the model's ability to capture and distinguish intricate details in the input data.
To quantitatively assess disorder, we compute the variance of the single-body correlators $\langle Z_i \rangle$ across all sites and summed over all time steps:
\begin{equation}
\mathrm{Variance} = \sum_{t} \frac{1}{N} \sum_{i=1}^N \left( \langle Z_i(t) \rangle - \overline{\langle Z(t) \rangle} \right)^2,
\end{equation}
where $\overline{\langle Z(t) \rangle}$ is the average over all sites at time $t$. Fig.(\ref{perf_heatmap}b) shows the heatmap of this variance as a function of $\Omega$ and interatomic distance. Importantly, the regions of highest variance coincide with the regions of optimal model performance in Fig.(\ref{perf_heatmap}a). This strong correlation supports our conclusion that maximizing disorder within an appropriate range yields more expressive quantum embeddings and thus enhances learning outcomes. This analysis was performed using the TDVP one-site method. A similar analysis for the TDVP two-site method is shown in Fig.(\ref{qelm_pic}b), where the highlighted regions indicate the optimal regime for model performance.

To further substantiate the relationship between disorder and model performance, we also compute the Edwards-Anderson (EA) order parameter, a well-established metric for quantifying the degree of order in disordered systems such as spin glasses \cite{ea}. In our context, a lower EA parameter indicates higher dynamical disorder in the quantum state. The EA order parameter, summed over time, is defined as:
\begin{equation}
q =  \sum_{t}  \sum_{i=1}^N \frac{1}{N}\langle Z_i(t) \rangle^2.
\end{equation}
Fig.(\ref{perf_heatmap}c) presents a heatmap of the EA parameter as a function of $\Omega$ and interatomic distance. The region where the EA parameter reaches its minimum aligns with the region of maximal model accuracy and highest variance in single-body correlators, reinforcing our conclusion that increased disorder in the quantum state leads to improved ML performance.
The underlying physical picture is that all three terms in the Hamiltonian, the Rabi drive $\Omega$, pairwise interaction $V_{jk}$, and inhomogeneous detuning must be in competition to generate rich, nontrivial quantum dynamics. For intermediate values of $\Omega$, when the distance between atoms is small, the interaction term $V_{jk}$ dominates, suppressing disorder and leading to lower accuracy. As the distance increases, the strength of $V_{jk}$ decreases, allowing the disorder term to become more significant and resulting in increased accuracy. However, with further increases in distance, the pairwise interaction becomes negligible compared to the other terms, and accuracy decreases once again.
Another important factor is that, for disorder to have a significant effect, all atoms must be free to transition between the ground and excited states, rather than being blockaded. This explains why the optimal distance is found to be greater than the blockade radius. The blockade radius $R_b$ is defined as the distance at which the Rydberg interaction energy equals the Rabi frequency and is given by $R_b = \left(\frac{C}{\Omega}\right)^{1/6}$, where $C$ is the interaction coefficient. For the case where $\Omega = 2\pi$, $R_b$ is calculated to be approximately $9.75~\mu\text{m}$, while optimal performance is observed at slightly larger distances.
In summary, our results demonstrate that disorder, tuned by the interplay of Hamiltonian parameters, plays a pivotal role in enabling expressive and robust quantum embeddings. This disorder driven diversity is essential for achieving high model accuracy in quantum inspired ELMs, even in the absence of significant entanglement.

\subsection{Scaling Analysis}
\begin{figure}
\centering \includegraphics[scale=0.45]{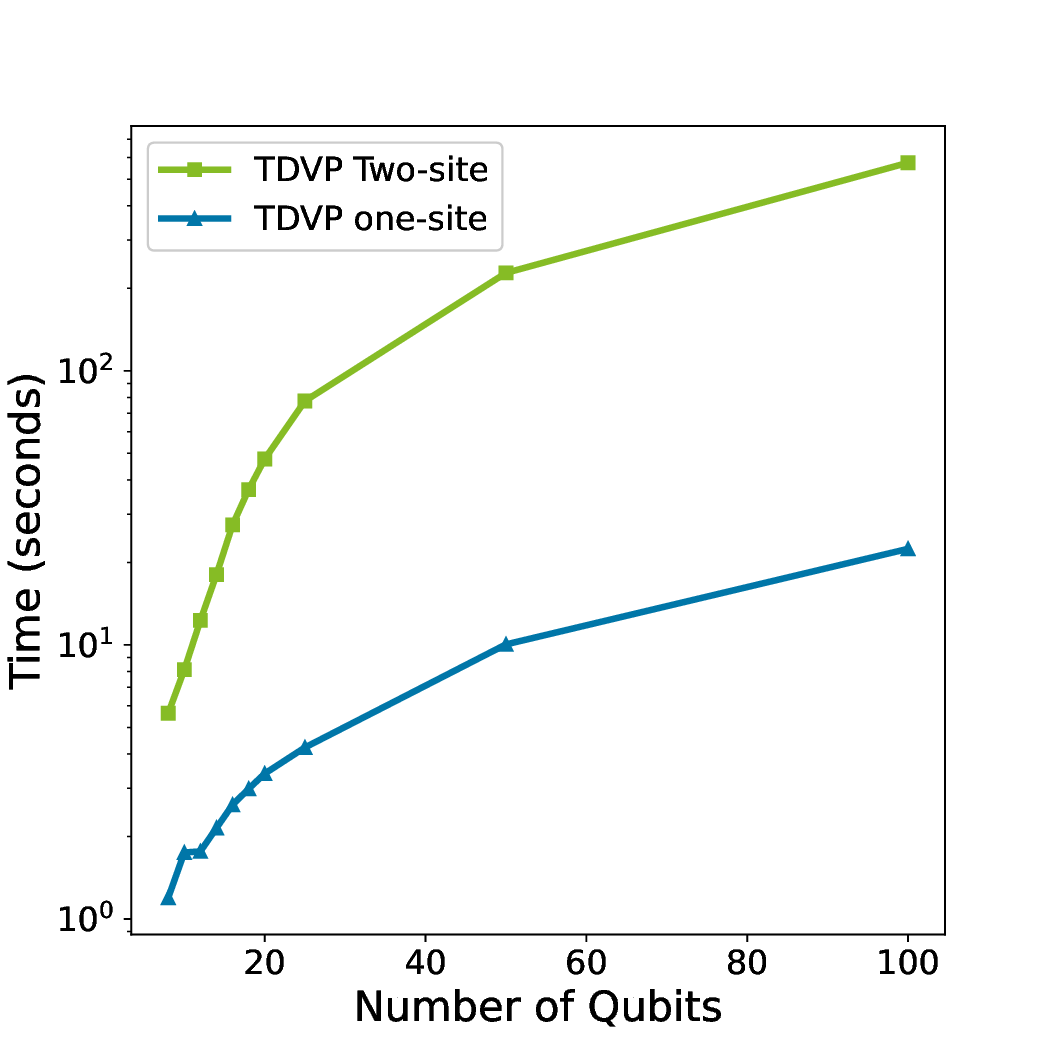} \caption{\textbf{Scaling analysis of different simulation methods to calculate the quantum embeddings in the QRC method}. Specifically, time complexity (in seconds) was taken for data embedding computation using the TDVP-two site, and TDVP-one site with increasing number of qubits on a normal laptop (AMD Ryzen 7 PRO 7730U with Radeon Graphics, 16.0 GB RAM). The experiments were conducted using the parameters $\Omega = 2\pi$ and $\phi = 0$. The ground state, where all qubits are in the ``up" state, was evolved using the Rydberg Hamiltonian up to $T = 4 \, \mathrm{\mu s}$ with time steps of $0.5 \, \mathrm{\mu s}$. For TDVP two-site, we have chosen the maximum bond dimension to be 100, and the bond dimension was fixed during the time evolution using the one-site TDVP. The y-axis is presented in logarithmic scale.}
\label{fig:time_computation}
\end{figure}

We begin our scaling analysis by examining the computational time complexity associated with generating quantum embeddings for single data points in the MNIST dataset. Specifically, we compare two TN simulation methods for evolving the Rydberg Hamiltonian: the TDVP two site method, with a maximum bond dimension set to 100, and the TDVP one site method. For this analysis, the interatomic distance is fixed at $d = 11~\mu\mathrm{m}$ and the Rabi frequency at $\Omega = 2\pi$. The computation time required for embedding generation as a function of the number of qubits is shown in Fig.(\ref{fig:time_computation}). 

Both TDVP methods exhibit a steady and manageable increase in computation time as the number of qubits grows, in stark contrast to the exponential scaling observed with exact diagonalization. Among the two, the TDVP one site method is notably more efficient, making it particularly well suited for practical applications involving larger quantum systems. This favorable scaling highlights the advantage of TN approaches for simulating quantum dynamics and generating quantum embeddings in ML contexts.

\begin{figure}
\centering \includegraphics[scale=0.35]{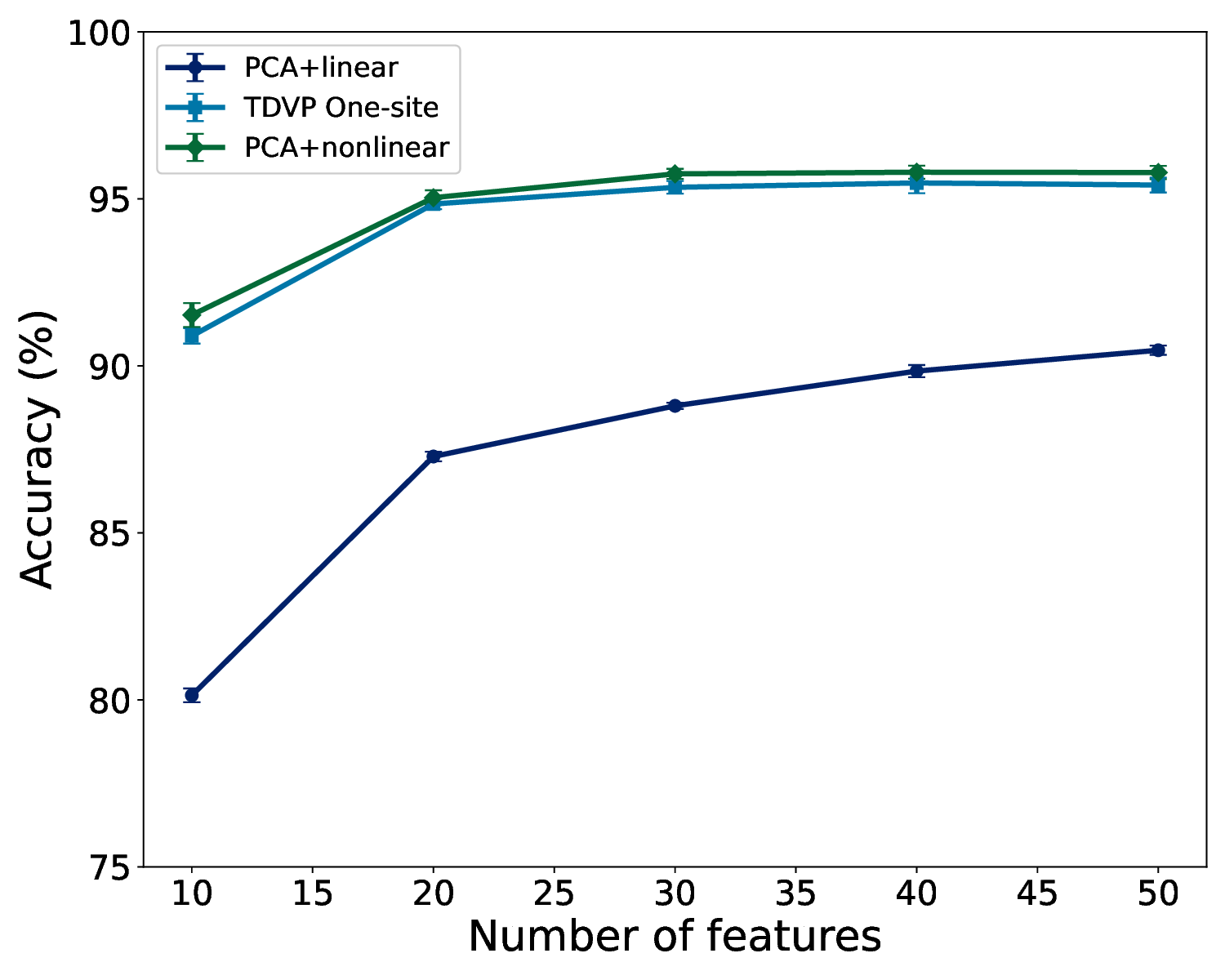} \caption{\textbf{Model accuracy comparison between the quantum-inspired QRC methods, and classical ML algorithms}.  This graph was generated using whole dataset and k fold cross validation using $k=5$. } \label{fig:accuracy_comparison} \end{figure}
Having established the computational efficiency of the TN methods, we next assess the impact of increasing the number of features on model accuracy. Using the TDVP one-site method, we compare the performance of the quantum-inspired QELM model to two classical baselines, a linear model trained on PCA-reduced data (PCA+linear) and an NN trained on PCA-reduced data (PCA+nonlinear). The results, presented in Fig.(\ref{fig:accuracy_comparison}), demonstrate that the QELM method consistently outperforms the linear model with classical features. As the number of qubits increases, the accuracy of the QELM model also increases before reaching a saturation point, reflecting the enhanced representational power provided by richer quantum embeddings. Importantly, the accuracy achieved by the QELM embedding using the TDVP one site method matches that of the nonlinear NN model with classical features, within the margin of error. This suggests that the QELM approach introduces sufficient nonlinearity through quantum dynamics, enabling a simple linear model to achieve high performance comparable to more complex classical NN. The error bars in Fig.(\ref{fig:accuracy_comparison}), obtained from $k$-fold cross-validation with $k=5$, are relatively small, indicating the consistency and reliability of these results.
It is important to note that while the TDVP one site method offers superior efficiency and scalability, TN simulations may not fully capture the true quantum dynamics for very large systems.  Nonetheless, our findings show that the embeddings generated by the TDVP one site method are sufficient to enable robust and optimal ML performance.
\section{Conclusion and Outlook}
In this work, we investigated QELM through the lens of tensor network algorithms, presenting a quantum-inspired approach to enhance classical machine learning tasks. By encoding data features as site-dependent detunings within a Rydberg Hamiltonian and simulating the system’s time evolution, we leveraged tensor network techniques to efficiently extract high-quality quantum embeddings. Our results demonstrate that this approach enables the simulation of quantum systems with a large number of qubits on classical hardware, providing expressive embeddings. A notable advantage of the tensor network-based method is its ability to control and manage entanglement, thereby mitigating the exponential concentration problem that can hinder learning in QELM. Our numerical experiments on the MNIST dataset highlight that optimal model performance is achieved by tuning Hamiltonian parameters to maximize disorder while maintaining moderate entanglement in the quantum states.

Looking ahead, it will be important to explore how this algorithm can be applied to more complex or real-world datasets, assessing its scalability and effectiveness in practical scenarios. Additionally, further research is needed to delineate the boundaries of tensor network methods in simulating quantum dynamics for various QML applications. In particular, investigating other datasets and scenarios where the exact simulation of quantum dynamics and the associated rapid growth of entanglement plays a critical role in machine learning performance would be valuable \cite{en1, en2}. We hope these findings encourage continued exploration of tensor network approaches for advancing both quantum-inspired and quantum machine learning.

\bigskip
\noindent
\textbf{Acknowledgements}
We would like to thank Milan Kornja\v ca, Sheng-Tao Wang, Mekena Mcgrew, and Scott Buchholz for their feedback and discussion. We are also grateful to Audrey Adib-Yazdi for her assistance in creating the graphics that accompany this article.

\bigskip
\noindent
\textbf{Disclaimer}
As used in this publication, “Deloitte” means Deloitte Consulting LLP, a subsidiary of Deloitte LLP. Please see www.deloitte.com/us/about for a detailed description of our legal structure. Certain services may not be available to attest clients under the rules and regulations of public accounting. This publication contains general information only and Deloitte is not, by means of this publication, rendering accounting, business, financial, investment, legal, tax, or other professional advice or services. This publication is not a substitute for such professional advice or services, nor should it be used as a basis for any decision or action that may affect your business. Before making any decision or taking any action that may affect your business, you should consult a qualified professional advisor.

Deloitte shall not be responsible for any loss sustained by any person who relies on this publication. 

Copyright \copyright\; 2025 Deloitte Development LLC. All rights reserved..

\bibliography{apssamp}% Produces the bibliography via BibTeX.
\clearpage
\onecolumngrid
\setcounter{section}{0}
\setcounter{subsection}{0}
\setcounter{subsubsection}{0}
\setcounter{figure}{0}
\setcounter{table}{0}
\setcounter{equation}{0}
\setcounter{page}{1} % Optional: resets page numbering

\begin{center}
    \textbf{Supplementary Material for ``Harnessing Physics-Inspired Dynamics for Robust and Scalable Quantum Extreme Learning Machines''}
\end{center}

\twocolumngrid
\section{Details of parameters for machine learning models}
\begin{itemize} \item {\bf Linear fitting of the original data}: A simple linear model was trained on the PCA-reduced data, consisting of a single dense layer with 10 units and L1 regularization. The optimizer used was Adam, and the loss function was Sparse Categorical Cross Entropy. \item {\bf Non-linear fitting of the original data}: A 4-layer feedforward neural network (NN) model with two hidden layers and L1 regularization was trained on the PCA-reduced data. The architecture includes two dense layers with 100 units each (Rectified Linear Unit (ReLU) activation) and a dense output layer with 10 units (softmax activation). The optimizer used was Adam, and the loss function was Sparse Categorical Crossentropy. \item {\bf Linear fitting of quantum embeddings}: A linear model with L1 regularization was trained on the quantum embeddings. The model consisted of a single dense layer with 10 units and an L1 regularizer. The optimizer used was Adam, and the loss function was Sparse Categorical Cross Entropy. \end{itemize}

\section{Effect of $V_{jk}$ Threshold}

\begin{figure}[t]
    \centering
    \includegraphics[scale=0.3]{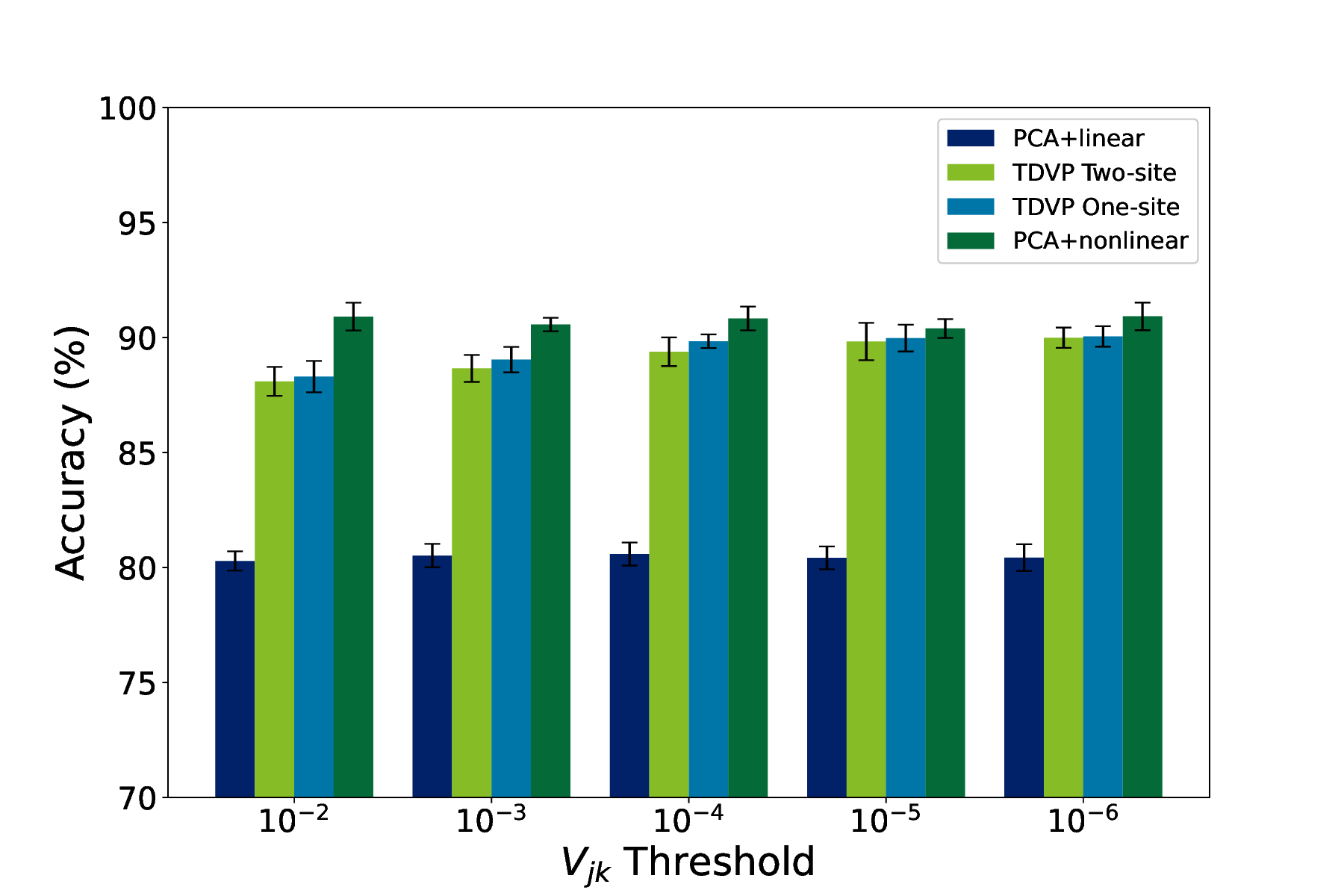}
    \caption{\textbf{Quantum inspired QELM performance for 10 qubits (10 features) with varying $V_{jk}$ threshold for Rydberg Hamiltonian}. This result was compared against the quantum embedding generated by the TDVP one site and TDVP two site method. For this study, we also compared the quantum-inspired results against results from classical models (linear and NN) with classical features.}
    \label{v_jk_threshold}
\end{figure}

In section \ref{TN_method}, the interaction term $V_{jk}$ was discussed, which becomes increasingly negligible for the farthest neighbors. These terms can be disregarded if $V_{jk}$ is below a certain threshold value. In this section, we will examine how this threshold impacts model accuracy. To this end, we have plotted a graph of model accuracy across various models against the $V_{jk}$ threshold, as shown in Fig.(\ref{v_jk_threshold}).
By increasing the threshold, more neighbors are incorporated, as well as additional features in the embedding. Consequently, the accuracy of the model increases as the threshold decreases. It is important to note that the time required for performing the time evolution also increases with the inclusion of added long-range interactions. To balance overall performance, we have chosen the $V_{jk}$ threshold to be $10^{-4}$ throughout the investigations.

\begin{figure*}
    \centering
    \includegraphics[scale=0.4]{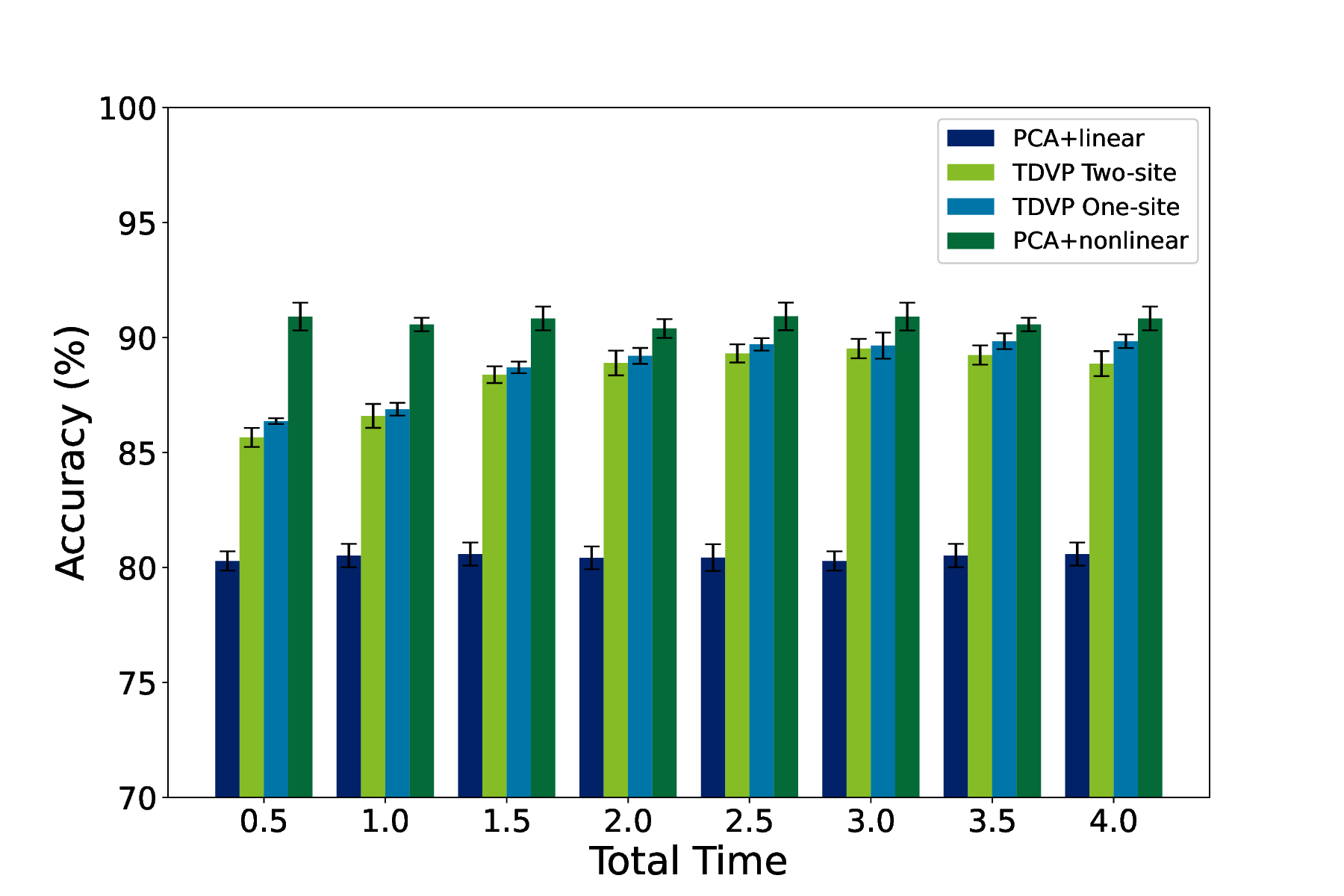}
    \caption{\textbf{Quantum-inspired QELM performance for 10 qubits (10 features) with varying total time for time evolution}. This result was compared against the quantum embedding generated by the TDVP one site and TDVP two site method. For this study, we also compared the quantum-inspired results against results from classical models (linear and NN) with classical features.}
    \label{accu_t}
\end{figure*}
\begin{figure*}
    \centering
    \includegraphics[scale=0.28]{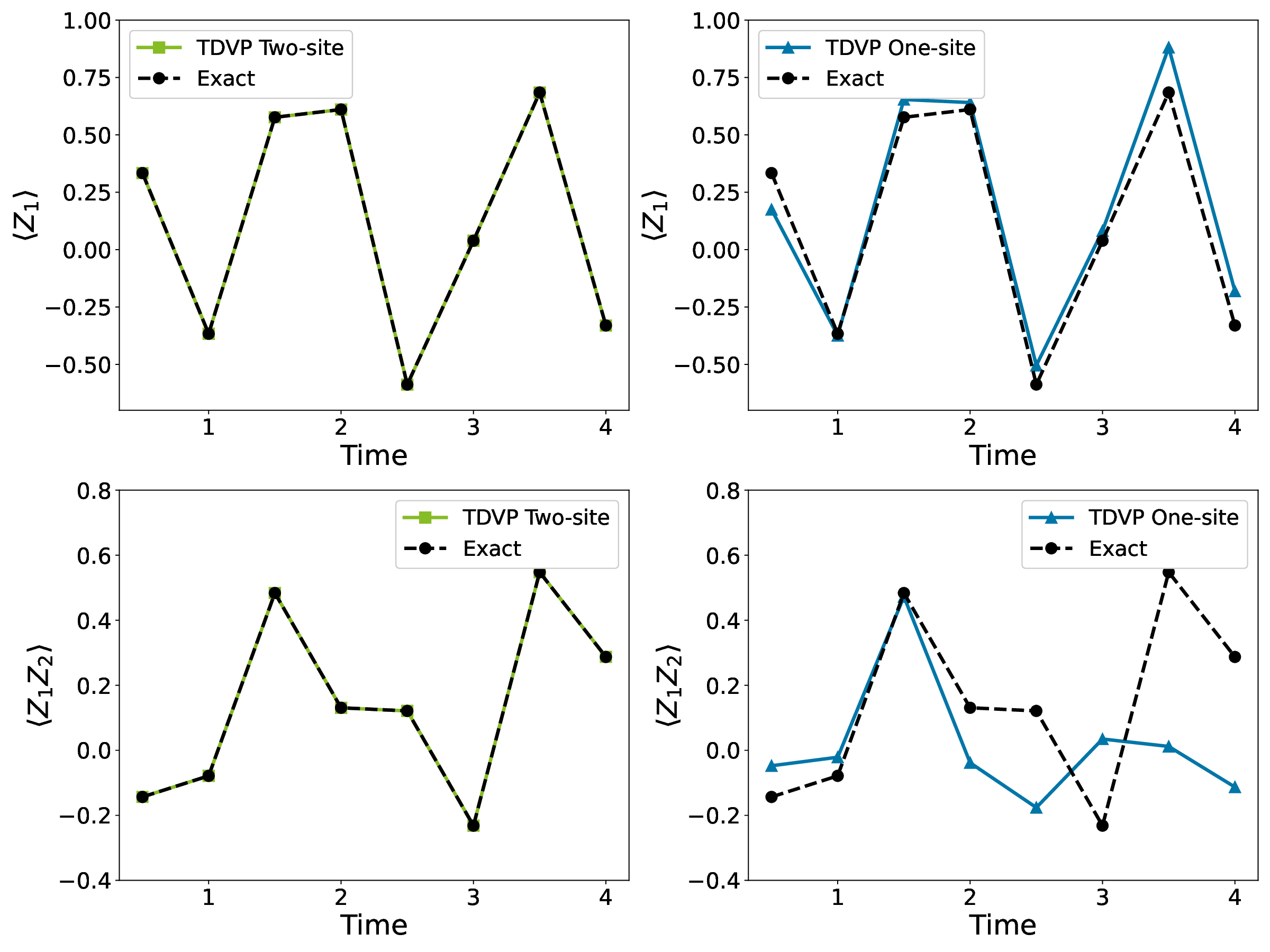}
    \caption{\textbf{Accuracy of different TN methods in calculating the time-evolution dynamics of 1D chain of Rydberg atoms}. Here we give an example of computing the expectation values associated with the dynamics of the first and second qubit. The one-body  $\langle Z_1 \rangle$ and the two-body  $\langle Z_1 Z_2 \rangle$ expectation values at different time steps were calculated using the TDVP two-site, and TDVP one-site. The results were compared against exact diagonalization. The maximum bond dimension was set at 100 for the TDVP two-site. The experiments were conducted using 8 qubits, and the parameters of the Rydberg Hamiltonian are: $\Omega = 2.2\pi$, $\Delta = 2.4\pi$, and $\phi = 0$.}
    \label{fig:comparison}
\end{figure*}
\section{Total Time for Time Evolution}
To investigate the effect of total time for time evolution, data embeddings were constructed at intervals of every 0.5 $\mu$s. Using these embeddings, we evaluated the model performance at each time step for all four machine learning models: TDVP one-site, TDVP two-site, and classical linear and nonlinear models, as illustrated in Fig.(\ref{accu_t}). It can be observed from Fig. \ref{accu_t} that the accuracy of the models increases as the total time progresses. This enhancement in model performance is primarily attributed to the augmentation of features within the data, which provides a richer representation for the machine learning models to leverage. As the total time increases, more intricate patterns and relationships within the dataset are captured, thereby improving the predictive capabilities of the models.

However, it is noteworthy that beyond the range of 2.5 to 3 $\mu$s, the accuracy reaches a plateau. This saturation suggests that the additional features incorporated beyond this time frame do not contribute any novel information to the models. Consequently, these features are deemed redundant, as they fail to enhance the model's ability to discern new patterns or improve its predictive accuracy. This observation underscores the importance of identifying an optimal feature set that balances complexity with meaningful information, thereby avoiding unnecessary computational overhead.

\section{Analysis of accuracy of correlators}
To begin with, we examined the accuracy of the correlators obtained using different methods. Specifically, we considered the one-body correlator $\langle Z_1 \rangle$ and the two-body correlator $\langle Z_1 Z_2 \rangle$ versus evolution time. Although the correlators derived from the tensor network algorithms may not be exact, their temporal trends closely resemble the exact results. This observation is illustrated in Fig.(\ref{fig:comparison}), where we compare the correlators (for an 8-qubit system) of data embeddings using the exact diagonalization method versus the TDVP one-site, and TDVP two-site methods. The TDVP two-site method, in particular, achieves a high degree of accuracy. It is important to note that for the purposes of QELM, the exact accuracy of the correlator values is not critical; the embeddings can still yield the appropriate model accuracy.

\end{document}